# Interaction of Shear-Horizontal Acoustic and Plasma Waves in Hexagonal Piezoelectric Semiconductors

Qingguo Xia[a], Jianke Du[a] and Jiashi Yang[b]

[a]Smart Materials and Advanced Structures Laboratory, School of Mechanical Engineering and Mechanics, Ningbo University, Ningbo, Zhejiang 315211, China

[b]Department of Mechanical and Materials Engineering, University of Nebraska-Lincoln, Lincoln, NE 68588-0526, USA

**Abstract**

We study coupled acoustic and plasma waves in piezoelectric semiconductor crystals of hexagonal symmetry. We focus on the so-called shear-horizontal or antiplane motions with one mechanical displacement. A set of two-dimensional equations is reduced from the three-dimensional equations. Since the material is effectively isotropic in the two-dimensional plane under consideration, the equations are relatively simple. Dispersion curves of coupled elastic and acoustic waves are obtained analytically and examined numerically along with the effects of some parameters.

## 1. Introduction

Some semiconductor crystals have electromechanical interactions such as piezoelectric or electrostrictive couplings. In these materials, the distribution and motion of mobile charges are affected by mechanical fields through the accompanying electric fields. In the second half of the last century, there were attempts to make devices based on these couplings. These early studies were reviewed in [1,2]. Since around the beginning of the present century, because of the capability of manufacturing new piezoelectric semiconductor structures [3], many new piezoelectric semiconductor devices have been developed, forming new research areas called piezotronics and piezophototronics [4,5]. This has attracted broad research interest with rapidly growing results, e.g., [6-18]. In most of these recent studies, the inertia of the charge carriers was usually neglected as an approximation. While this approximation is valid in many applications, it is well known that charge carriers such as electrons and holes have effective mass [19]. The inertia of charge carriers is necessary in the study of, e.g., the propagation of disturbances in charge carriers called plasma waves [20-23]. In this paper, we study the interaction of plasma waves with elastic waves in hexagonal piezoelectric semiconductors. We focus on the so-called shear-horizontal or anti-plane motions [18].

## 2. Governing Equations

We consider n-type semiconductors only. The equations below are based on the multi-continuum model in [24,25] consisting of four constituents of the lattice, bound charge, impurity and electron continua. The lattice and impurity continua move together. Let $\rho$ be the mass density of the lattice and impurity together. $\sigma^i$ is the charge density of the impurity. The bound-charge continuum is massless and can displace with respect to the lattice by a small displacement for describing electric polarization. The electron fluid can flow through the lattice. Its mass and charge densities are $\rho^e$ and $\sigma^e$ which are related by the charge-to-mass ratio of electrons.

We use the spatial description. All fields are functions of **y** and *t*. The velocity fields of the lattice continuum and the electron fluid are **v** and $\mathbf{v}^e$, respectively. The material time derivatives following the lattice and electron fluid are denoted by

$$\frac{d}{dt} = \frac{\partial}{\partial t} + \mathbf{v} \cdot \nabla, \quad \frac{d^e}{dt} = \frac{\partial}{\partial t} + \mathbf{v}^e \cdot \nabla, \quad \nabla = \mathbf{e}_i \frac{\partial}{\partial y_i}. \tag{2.1}$$

We have the following conservation of mass for the lattice, impurity and bound charge together:

$$\frac{\partial \rho}{\partial t} + \nabla \cdot (\rho \mathbf{v}) = 0. \tag{2.2}$$

The conservation of charge for the electron fluid is

$$\frac{\partial \sigma^e}{\partial t} + \nabla \cdot \mathbf{J}^e = 0, \tag{2.3}$$

where $\mathbf{J}^e$ is the electron current density given by

$$\mathbf{J}^e = \sigma^e \mathbf{v}^e. \tag{2.4}$$

In terms of the number density of electrons per unit volume, *n*, and the elementary charge, *q*, we have

$$\sigma^e = -qn, \quad \mathbf{J}^e = -qn\mathbf{v}^e. \tag{2.5}$$

Then (2.3) can be written as

$$-q\frac{\partial n}{\partial t} + \nabla \cdot \mathbf{J}^e = 0. \tag{2.6}$$

The quasistatic electric field **E**, electric displacement **D**, electric polarization **P** and the electrostatic potential $\varphi$ satisfy

$$\nabla \times \mathbf{E} = 0, \quad \mathbf{E} = -\nabla \varphi, \tag{2.7}$$

$$\nabla \cdot \mathbf{D} = \sigma^e + \sigma^i, \quad \mathbf{D} = \varepsilon_0 \mathbf{E} + \mathbf{P}. \tag{2.8}$$

For an n-type semiconductor, we can write

$$\sigma^i = qN_D^+, \tag{2.9}$$

where $N_D^+$ is the number density of ionized donors. Then the charge equation in (2.8) becomes

$$\nabla \cdot \mathbf{D} = q(N_D^+ - n). \tag{2.10}$$

The linear momentum equation for the lattice, impurity and bound charge together is

$$\nabla \cdot \boldsymbol{\tau} + \mathbf{P} \cdot \nabla \mathbf{E} + \sigma^i \mathbf{E} - \sigma^e \mathbf{E}^e = \rho \frac{d\mathbf{v}}{dt} \tag{2.11}$$

where $\boldsymbol{\tau}$ is the stress tensor in the combined continuum of the lattice, impurity and bound charge. $\mathbf{E}^e$ is an effective electric field representing the interaction between the lattice continuum and the electron fluid. In view of the linearization to be made throughout the rest of this paper, we drop the nonlinear term of $\mathbf{P} \cdot \nabla \mathbf{E}$ and simplify (2.11) to

$$\nabla \cdot \boldsymbol{\tau} + \sigma^i \mathbf{E} - \sigma^e \mathbf{E}^e = \rho \frac{d\mathbf{v}}{dt}. \tag{2.12}$$

The linear momentum equation for the electron fluid alone is

$$-\nabla p^e + \sigma^e \left( \mathbf{E} + \mathbf{E}^e \right) = \rho^e \frac{d^e \mathbf{v}^e}{dt}, \tag{2.13}$$

where $p^e$ is the pressure field in the electron fluid. In addition to the above field equations, we have the following constitutive relations of piezoelectric crystals:

$$\tau_{ij} = c_{ijkl} u_{k,l} + e_{kij} \varphi_{,k}, \quad D_i = e_{ikl} u_{k,l} - \varepsilon_{ik} \varphi_{,k}, \tag{2.14}$$

where **u** is the displacement vector of the lattice from its reference state to its present state. The pressure field of the electron fluid is related to the electron number density $n$ through [20]

$$p^e = kTn, \tag{2.15}$$

where $T$ is the absolute temperature and $k$ Boltzmann's constant. For the interaction between the electron fluid and the lattice, we take

$$\mathbf{E}^e = [\boldsymbol{\mu}^e]^{-1} (\mathbf{v}^e - \mathbf{v}), \tag{2.16}$$

where $[\boldsymbol{\mu}^e]$ is the electron mobility tensor or matrix. Then (2.12) and (2.13) become

$$\nabla \cdot \boldsymbol{\tau} + \sigma^i \mathbf{E} - \sigma^e [\boldsymbol{\mu}^e]^{-1} (\mathbf{v}^e - \mathbf{v}) = \rho \frac{d\mathbf{v}}{dt}, \tag{2.17}$$

$$-\nabla p^e + \sigma^e \mathbf{E} + \sigma^e [\boldsymbol{\mu}^e]^{-1} (\mathbf{v}^e - \mathbf{v}) = \rho^e \frac{d^e \mathbf{v}^e}{dt}. \tag{2.18}$$

With the use of (2.5), we can write (2.17) and (2.18) as

$$\nabla \cdot \boldsymbol{\tau} + \sigma^i \mathbf{E} + qn [\boldsymbol{\mu}^e]^{-1} (\mathbf{v}^e - \mathbf{v}) = \rho \frac{d\mathbf{v}}{dt}, \tag{2.19}$$

$$-\nabla p^e - qn \mathbf{E} - qn [\boldsymbol{\mu}^e]^{-1} (\mathbf{v}^e - \mathbf{v}) = \rho^e \frac{d^e \mathbf{v}^e}{dt}. \tag{2.20}$$

We assume a reference state with

$$n = n_0 = N_D^+. \tag{2.21}$$

The system is assumed to be in small-amplitude vibrations around the reference state under small disturbances. We write

$$n = n_0 + \tilde{n}, \tag{2.22}$$

where $\tilde{n}$ is small. Then (2.4) and (2.10) take the following form:

$$\mathbf{J}^e = \sigma^e \mathbf{v}^e = -qn \mathbf{v}^e \cong -qn_0 \mathbf{v}^e \tag{2.23}$$

$$\nabla \cdot \mathbf{D} = q(-\tilde{n}). \tag{2.24}$$

For small velocity fields and small velocity gradients, we approximate the material derivatives in (2.1) by partial derivatives. Then, for small electric fields, (2.19) and (2.20) are approximated by

$$\nabla \cdot \boldsymbol{\tau} + qn_0 \mathbf{E} + qn_0 [\boldsymbol{\mu}^e]^{-1} (\mathbf{v}^e - \mathbf{v}) = \rho^0 \frac{\partial \mathbf{v}}{\partial t}, \tag{2.25}$$

$$-\nabla p^e - qn_0 \mathbf{E} - qn_0 [\boldsymbol{\mu}^e]^{-1} (\mathbf{v}^e - \mathbf{v}) = \rho^{e0} \frac{\partial \mathbf{v}^e}{\partial t}, \tag{2.26}$$

where $\rho^0$ and $\rho^{e0}$, the reference mass densities of the lattice and electron fluid, have been used.

In summary, we have the conservation of charge in (2.6), the charge equation of electrostatics in (2.24), and the linear momentum equations in (2.25) and (2.26) which can be written as four linear equations for $\tilde{n}$, $\varphi$, **u** and $\mathbf{v}^e$ using (2.14), (2.15), (2.23) and $\mathbf{v} = \partial \mathbf{u}/\partial t$.

### 3. Shear-Horizontal Motions of Hexagonal Crystals

For hexagonal crystals such as ZnO, the relevant material matrices are

$$\begin{pmatrix} c_{11} & c_{12} & c_{13} & 0 & 0 & 0 \\ c_{21} & c_{11} & c_{13} & 0 & 0 & 0 \\ c_{31} & c_{31} & c_{33} & 0 & 0 & 0 \\ 0 & 0 & 0 & c_{44} & 0 & 0 \\ 0 & 0 & 0 & 0 & c_{44} & 0 \\ 0 & 0 & 0 & 0 & 0 & c_{66} \end{pmatrix}, \begin{pmatrix} \varepsilon_{11} & 0 & 0 \\ 0 & \varepsilon_{11} & 0 \\ 0 & 0 & \varepsilon_{33} \end{pmatrix}$$

$$\begin{pmatrix} 0 & 0 & 0 & 0 & e_{15} & 0 \\ 0 & 0 & 0 & e_{15} & 0 & 0 \\ e_{31} & e_{31} & e_{33} & 0 & 0 & 0 \end{pmatrix}, \begin{pmatrix} \mu_{11}^e & 0 & 0 \\ 0 & \mu_{11}^e & 0 \\ 0 & 0 & \mu_{33}^e \end{pmatrix}, \quad (3.1)$$

where $c_{66} = (c_{11}-c_{12})/2$. For shear-horizontal motions,

$$u_1 = u_2 = 0, \quad u_3 = u_3(x_1, x_2, t),$$
$$v_1^e = v_1^e(x_1, x_2, t), \quad v_2^e = v_2^e(x_1, x_2, t), \quad v_3^e = 0, \quad (3.2)$$
$$\varphi = \varphi(x_1, x_2, t), \quad \tilde{n} = \tilde{n}(x_1, x_2, t).$$

The relevant constitutive relations are

$$T_{31} = c_{44} u_{3,1} + e_{15} \varphi_{,1}, \quad T_{23} = c_{44} u_{3,2} + e_{15} \varphi_{,2}, \quad (3.3)$$
$$D_1 = e_{15} u_{3,1} - \varepsilon_{11} \varphi_{,1}, \quad D_2 = e_{15} u_{3,2} - \varepsilon_{11} \varphi_{,2}, \quad (3.4)$$
$$J_1^e = -q n_0 v_1^e, \quad J_2^e = -q n_0 v_2^e, \quad (3.5)$$
$$p^e = kT(n_0 + \tilde{n}). \quad (3.6)$$

(2.6) and (2.24) reduce to

$$-q \frac{\partial \tilde{n}}{\partial t} + J_{1,1}^e + J_{2,2}^e = 0, \quad (3.7)$$
$$D_{1,1} + D_{2,2} = q(-\tilde{n}). \quad (3.8)$$

As an approximation, we drop some of the terms in (25) and (26) so that they reduce to the linear version of the equations in [20]

$$\nabla \cdot \boldsymbol{\tau} = \rho^0 \frac{\partial \mathbf{v}}{\partial t}, \quad (3.9)$$

$$-\nabla p^e - q n_0 \mathbf{E} = \rho^{e0} \frac{\partial \mathbf{v}^e}{\partial t}. \quad (3.10)$$

Substituting (3.3)-(3.6) into (3.7)-(3.10), we obtain the following five equations for $u_3$, $\varphi$, $\tilde{n}$, $v_1^e$ and $v_2^e$:

$$c_{44}(u_{3,11}+u_{3,22}) + e_{15}(\varphi_{,11}+\varphi_{,22}) = \rho^0 \frac{\partial^2 u_3}{\partial t^2},$$

$$e_{15}(u_{3,11}+u_{3,22}) - \varepsilon_{11}(\varphi_{,11}+\varphi_{,22}) = q(-\tilde{n}),$$

$$q\frac{\partial \tilde{n}}{\partial t} + qn_0 v^e_{1,1} + qn_0 v^e_{2,2} = 0, \qquad (3.11)$$

$$-kT\tilde{n}_{,1} + qn_0 \varphi_{,1} = \rho^{e0} \frac{\partial v^e_1}{\partial t},$$

$$-kT\tilde{n}_{,2} + qn_0 \varphi_{,2} = \rho^{e0} \frac{\partial v^e_2}{\partial t},$$

which can be reduced to

$$c_{44}qn_0^2(u_{3,11}+u_{3,22}) + e_{15}kTn_0(\tilde{n}_{,11}+\tilde{n}_{,22}) = qn_0^2 \rho^0 \frac{\partial^2 u_3}{\partial t^2} + e_{15}\rho^{e0} \frac{\partial^2 \tilde{n}}{\partial t^2}, \qquad (3.12)$$

$$e_{15}qn_0^2(u_{3,11}+u_{3,22}) - \varepsilon_{11}kTn_0(\tilde{n}_{,11}+\tilde{n}_{,22}) = -\varepsilon_{11}\rho^{e0} \frac{\partial^2 \tilde{n}}{\partial t^2} - q^2 n_0^2 \tilde{n}.$$

## 4. Plane Waves

Consider plane waves propagating in the $x_1$ direction with

$$u_3 = A_1 \exp[i(\xi x_1 - \omega t)], \quad \tilde{n} = A_2 \exp[i(\xi x_1 - \omega t)]. \qquad (4.1)$$

Substituting (4.1) into (3.12), we obtain

$$(qn_0^2\rho^0\omega^2 - c_{44}qn_0^2\xi^2)A_1 + (e_{15}\rho^{e0}\omega^2 - n_0 e_{15}kT\xi^2)A_2 = 0,$$

$$-e_{15}qn_0^2\xi^2 A_1 + (n_0\varepsilon_{11}kT\xi^2 - \varepsilon_{11}\rho^{e0}\omega^2 + q^2 n_0^2)A_2 = 0. \qquad (4.2)$$

For nontrivial solutions, the determinant of the coefficient matrix of (4.2) has to vanish, which leads to

$$\frac{\omega^4}{\xi^2} - v_{Te}^2\omega^2 - \frac{\omega_{pe}^2\omega^2}{\xi^2} - c_s^2\omega^2 - \frac{e_{15}^2\omega^2}{\varepsilon_{11}\rho^0} + v_{Te}^2 c_s^2\xi^2 + c_s^2\omega_{pe}^2 + \frac{v_{Te}^2 e_{15}^2\xi^2}{\varepsilon_{11}\rho^0} = 0, \qquad (4.3)$$

where

$$c_s = \sqrt{\frac{c_{44}}{\rho^0}}, \quad v_{Te} = \sqrt{\frac{kTn_0}{\rho^{e0}}}, \quad \omega_{pe} = \sqrt{\frac{q^2 n_0^2}{\rho^{e0}\varepsilon_{11}}}. \qquad (4.4)$$

(4.3) can be written as

$$\left[\frac{\omega^2}{\xi^2} - \left(c_s^2 + \frac{e_{15}^2}{\varepsilon_{11}\rho^0}\right)\right](\omega^2 - v_{Te}^2\xi^2 - \omega_{pe}^2) = \frac{\omega_{pe}^2 e_{15}^2}{\varepsilon_{11}\rho^0}. \qquad (4.5)$$

When $e_{15}=0$, (4.5) decouples into separate elastic and plasma waves with

$$\frac{\omega^2}{\xi^2} - \left(c_s^2 + \frac{e_{15}^2}{\varepsilon_{11}\rho^0}\right) = 0,$$

$$\omega^2 - v_{Te}^2\xi^2 - \omega_{pe}^2 = 0. \qquad (4.6)$$

For a numerical example, consider ZnO with $\rho^0 = 5680$ kg/m$^3$, $n_0 = 10^{21}$ /m$^3$, $T = 300$ K, $e_{15} = -0.48$ C/m$^2$, $c_{44} = 4.247\times 10^{10}$ N/m$^2$, $\varepsilon_{11} = 8.55\varepsilon_0$, $\varepsilon_0 = 8.854\times 10^{-12}$ F/m, $q = 1.602\times 10^{-19}$ C, $\rho^{e0} = n_0 m_e$, $m_e = 0.014 m_0$, and $m_0 = 9.11 \times 10^{-31}$ Kg. Figure 1 shows the dispersion curves of uncoupled plasma and elastic waves from (4.6). Figure 2 shows the corresponding coupled waves from (4.5). Figure 3 shows the effect of $n_0$ on coupled waves. In Fig. 4, we artificially vary the value of $e_{15}$ alone while keeping the other materials constants unchanged.

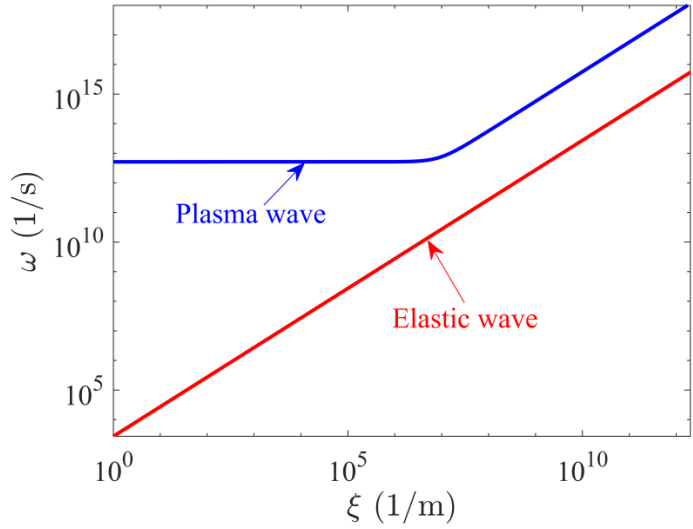

Fig. 1. Uncoupled plasma and elastic waves

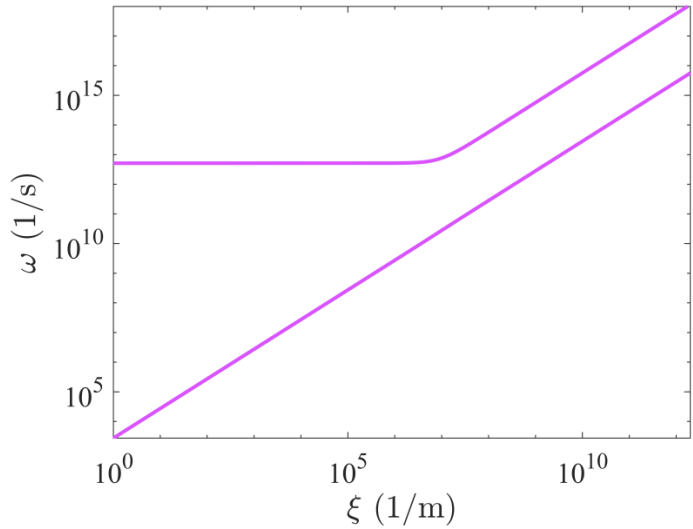

Fig. 2. Coupled plasma and elastic waves

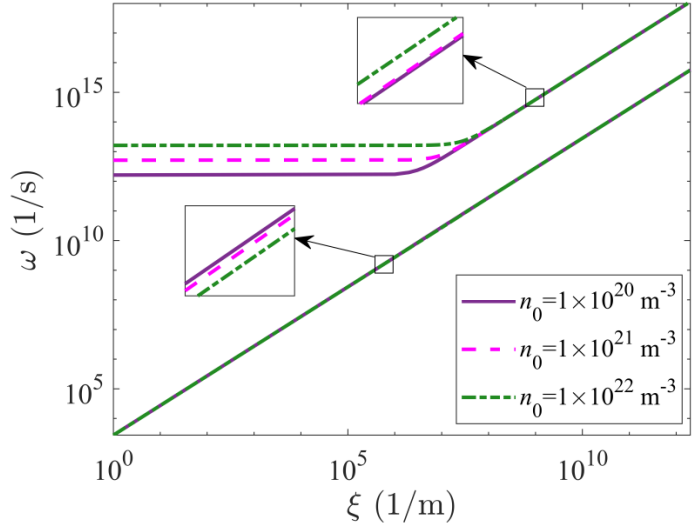
Fig. 3. Effects of $n_0$ on coupled waves

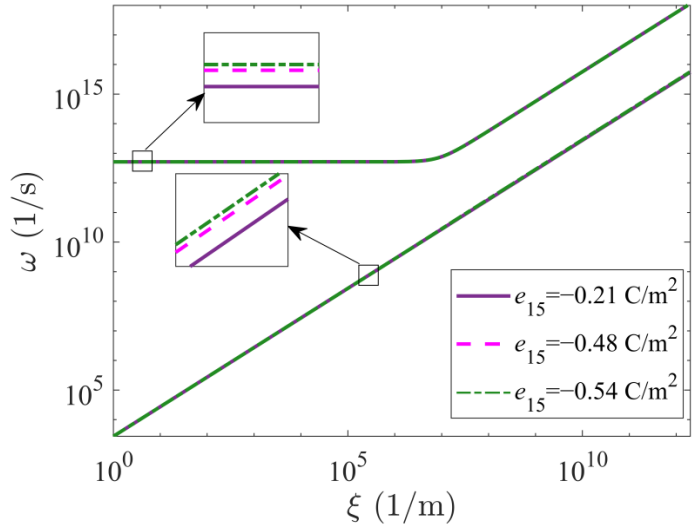
Fig. 4. Effects of $e_{15}$ on coupled waves

## 5. Conclusions

For piezoelectric semiconductors of hexagonal symmetry such as ZnO, the equations governing shear-horizontal motions are relatively simple mathematically. Solutions of the dispersion curves of coupled elastic and plasma waves are obtained, showing that the coupling relies on the piezoelectric effect of the material. Numerical results show that the dispersion curves are sensitive to the parameters of the material such as the doping level.